# Looking into the Future of Health-Care Services: Can Life-Like Agents Change the Future of Health-Care Services?

Mohammad Saleh Torkestani [a,*], Robert Davis [b], Abdolhossein Sarrafzadeh [c]

[a] *Visiting Scholar, Dept. of Computing, Unitec, New Zealand*
[b] *Head of Dept., Dept. of Management and Marketing, Unitec, New Zealand*
[c] *Head of Dept., Dept. of Computing, Unitec, New Zealand*

**Abstract**

Time constraints on doctor–patient interaction and restricted access to specialists under the managed care system led to increasingly referring to computers as a medical information source and a self-health-care management tool. However research show that less than 40% of information seekers indicated that online information helped them to make a decision about their health. Searching multiple web sites that need basic computer skills, lack of interaction and no face to face interaction in most search engines and some social issues, led us to develop a specialized life-like agent that would overcome mentioned problems.

## 1. Introduction

Enjoyment of the highest attainable standard of health is one of the fundamental rights of every human being (WHO statement) [34]. The increasing availability of computer-mediated knowledge and the advancement of information and communication technologies have altered the methods through which health care information is sought [3] [25] [30]. The Internet has had a significant impact on healthcare service and is a virtual medical library for an estimated 75–80% of users in developed countries [4] [5] [11]. On an average day, more than six million patients and their caregivers in the United States use the Internet to obtain health and medical information. This number exceeds the average daily number of 2.27 million Americans who make visits to physician offices [11] [18] [26]. Furthermore, not only patients but their caregivers want to get actively involved in the health-care management of their loved ones. In a research nearly 60% of people who identified themselves as caregivers use the Internet to find answers to their health-related questions [16].

This computer mediated environment has become, as Vargo and Lusch [32] argue, a fundamental hub where "people exchange to acquire the benefits of specialized competencies (knowledge and skills), or services."

## 2. Health Care Systems

During the past few decades computer-based automated systems have been integrated into many facets of peoples' lives. In health care, computer-based systems designed to provide health information and advice to patients and consumers are on the rise [27]. Some of these systems appear similar to a human health professional: they are interactive and designed to emulate the style of a human counselor [7]. They ask questions, provide tailored and appropriate responses to users, offer practical advice and educate by providing useful and up-to-date health information. Some computer-based systems are deliberately designed to display human qualities, such as empathy, humor and objectivity [8]. The goal is to have users respond as they would to a human health professional.

* Corresponding author. Tel.: +64-9-8154321.
  *E-mail address*: storkestani@Unitec.ac.nz or torkestani@gmail.com.

Studies of users of automated systems, in health and other spheres of human life, note that many users actually respond to them in anthropomorphic terms [6] [13]: users assign human characteristics such as emotions, feelings and other qualities to the inanimate computer system.

Electronic health information on the web is not limited to simple non-interactive informational sites. Among the web-based applications are diagnostic tools where patients, who suspect they have a medical issue, can enter their basic demographic information, a description of their symptoms, and a credit-card number to receive a preliminary diagnosis despite potential pitfalls of being misinformed [25] [28].

## 3. Artificial Intelligence and Health Care Systems

Nowadays, researchers are increasingly using information technology to examine new and innovative techniques to overcome the rapid surge in health care costs facing the community. Research undertaken in the past has shown that artificial intelligence (AI) tools and techniques can aid in the diagnosis of disease and the assessment of treatment outcomes. [1] Recent developments in information technology and AI tools, particularly in neural networks, fuzzy logic and support vector machines, have provided the necessary support to develop highly efficient automated diagnostic systems. Despite plenty of challenges, these new advances in AI tools hold much promise for solving medical and health-related problems.[2] [9] There have been a number of artificial intelligence (AI) tools developed over the past decade or so [10] [14]. Many of these have found their applications in medical and health-related areas. Commonly applied AI techniques can be listed as: Neural Networks, Fuzzy Logic, Support Vector Machines, Genetic Algorithms and Hybrid Systems. In addition to applications in medical diagnostic systems, AI techniques have been applied in many biomedical signal-processing tasks [19]. Neural network models have played a dominant role in the majority of AI-related applications in health and medicine. Many of these applications are for pattern recognition or classification. A typical classification application usually has a number of steps or procedures, as shown by the flow diagram (see Fig. 1). This involves feature extraction from the input data before feeding these features to the classifier for designing and developing automated classification models, and finally testing the models for generalization.

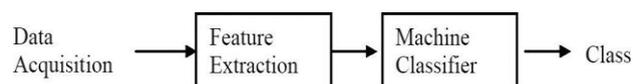

Figure 1.   Stages of a typical pattern recognition task [19]

## 4. Intelligent Agents and Life-Like Agents

Life-Like agents are a special kind of so-called intelligent software agents. Software agent technology that originated from distributed artificial intelligence is inherently interdisciplinary. Thus, the notion of agency is quite broadly used in the literature; it can be seen as a tool for analyzing systems, not an absolute characterization that divides the world into agents and non-agents [12] [33]. However, intelligent agents are commonly assumed to exhibit autonomous behavior determined by their:
- pro-activeness, which means taking the initiative to satisfy given design objectives and exhibiting goal-directed behavior
- Reactive or deliberative actions, which means perceiving the environment and using timely change management to meet given design objectives, and

- Social cooperation in groups with other agents and/or human users when needed.

What an intelligent agent in practice is supposed to do depends on the concrete application domain and on the potential for a solution for a particular problem. Today, intelligent agents are deployed in different settings, such as industrial control, Internet searching, personal assistance, network management, games and software distribution.

An intelligent agent for the Internet is commonly called information agent. It is autonomous, computational software entity (an intelligent agent) that has access to one or more heterogeneous and geographically distributed information sources, and which pro-actively acquires, mediates, and maintains relevant information on behalf of users or other agents, preferably just-in-time. Thus, the purpose of an information agent is supposed to satisfy one or more of the following requirements:

- Information acquisition and management. It is capable of providing transparent access to one or many different information sources. Furthermore, it retrieves, extracts, analyses, and filters data, monitors sources, and updates relevant information on behalf of its users or other agents. In general, the acquisition of information encompasses a broad range of scenarios, including advanced information retrieval in databases and also the purchase of relevant information from providers in electronic marketplaces.
- Information synthesis and presentation. The agent is able to fuse heterogeneous data and provide unified, multi-dimensional views on information relevant to the user.
- Intelligent user assistance. The agent can dynamically adapt to changes in user preferences, information, and the network environment as well. It provides intelligent, interactive assistance for common users, supporting their information-based business on the Internet. In this context, the utilization of intelligent user interfaces like believable, life-like characters can significantly increase not only the awareness of the user of their personal information agent but also the way information is interactively dealt with.

Many (systems of) information agents have been developed or are currently under development in academic and commercial research labs, but they still have to wait to make it out to the real world of Internet users. However, the ambitious goal of satisfying all of the requirements mentioned above appears close to accomplishment in the next ten years.

Information agents may be categorized into several different classes according to one or more of the following features [15]:

- Non-cooperative or cooperative information agents depending on the ability of the agents to cooperate with each other for the execution of their tasks. Several protocols and methods are available for achieving cooperation among autonomous information agents in different scenarios, like hierarchical task delegation, contracting, and decentralized negotiation.
- Adaptive information agents are able to adapt themselves to changes in networks and information environments. Examples of such kinds of agents are learning personal assistants on the Web.
- Rational information agents behave in an economically rational way, based on utilitarian decision making. As self-interested agents they aim to increase their own benefits. One main application domain of such agents is automated trading and electronic commerce on the Internet. Examples include the variety of 'shop-bots' and systems for agent-mediated auctions on the Web.
- Mobile information agents can travel autonomously through the Internet. Such agents may enable, for example, dynamic load balancing in large-scale networks, reduction of data transfer among information servers, applications, and migration of small business logic within medium-range corporate intranets on demand.

Representing and processing ontological knowledge and metadata, user profiles and natural language input, translation of data formats as well as machine learning techniques enables an information agent to acquire and maintain knowledge about itself and its user, the network, and the information environment.

In a conclusion, Life-like agents are software agents with personalities. They are versatile and employ friendly front ends to communicate with users. They are animated computer representations of humanlike movements and behaviors in a computer-generated three-dimensional world. Some life-like agents can speak and exhibit behaviors such as gestures and facial expressions. They can be fully automated to act like robots. The purpose of them is to introduce believable emotions so that the agents gain credibility with users [31].

## 5. Our Life-Like Agent: Dr. Eve!

In our research we will use an advanced life-like agent named "Dr. Eve" that is a combination of a life-like agent developed by one of the authors [22] [23] [24] and another open-source Life-Like agent as Computer-Mediated Communication tool. Dr. Eve is the beginning of a big change in the way everyone will use and interact with computers as a health care advisor. Dr. Eve's Artificial Engine uses several algorithms for its Artificial Intelligence. One of these layers uses AIML programming language. AIML stands for Artificial Intelligence Mark-up Language. It's an XML dialect for creating natural language applications, developed by Dr. Richard Wallace and a worldwide free software community [21]. AIML describes a class of data objects called AIML objects and partially describes the behavior of computer programs that process them. AIML objects are made up of units called topics and categories, which contain either, parsed or unparsed data. Parsed data is made up of characters, some of which form character data, and some of which form AIML elements. AIML elements encapsulate the stimulus-response knowledge contained in the document. Character data within these elements is sometimes parsed by an AIML interpreter, and sometimes left unparsed for later processing by a Responder.

Fig. 2 shows our proposed system architecture. The main structure of the system consists of a knowledge-based schema. This knowledge base is fed by two main resources. The first is direct feeding by designers using AIML programming language. The second source of Dr. Eve's knowledge base is trusted websites.

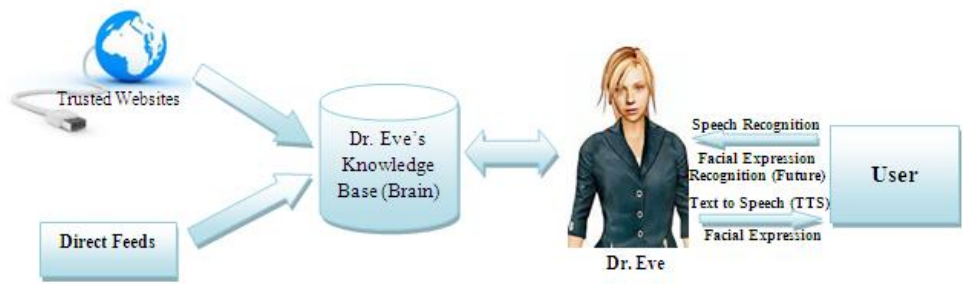

Figure 2. Architecture of proposed system.

As we mentioned above, many research have shown that healthcare consumers are concerned about getting low-quality health information on the Web [20] so government intervention is warranted. Examples are: The United Kingdom's "NHS Direct", Australia's "HealthInsite", Canada's "Canadian Health Network (CHN)", and United States of America's "HealthFinder" sites [29]. These websites are usually designed as a trusted source of health care information, and they could be the source of Dr. Eve's knowledge base as well.

Dr. Eve interacts with users using speech recognition module and Text to Speech (TTS) module. With speech recognition module Dr. Eve will try to understand all single word users say and will try to find the best answer to their question using her knowledge base. With text to speech module Dr. Eve talks to users as native female English speaker.

The face plays a significant role in social communication since it is a 'window' to human personality, emotions and thoughts. According to the psychological research conducted by Mehrabian [17], the nonverbal part is the most informative channel in social communication. As the lack of interaction and no face-to-face interaction in most former web based health care advisors was one of the main weaknesses of those systems, at the first stage Dr. Eve comes with a graphical facial interface that can express many of facial expressions of a real girl.

In the next stages we will try to add Facial Expression Recognition (FER) module to our system. Using this module, Dr. Eve could detect and recognize the facial expression of users and it could help her to adapt her interaction with each user according to his/her facial expression.

## 6. Conclusion

In this paper we reviewed some important trends in health care services and presented a new system based on Artificial Intelligence and Life-Like Intelligent Agents. This system is a new idea in patient - health care systems interaction and we believe that it could change the future of health care services. At this stage Dr. Eve is supported by some basic modules like Text To speech (TTS), Facial Expression and Speech recognition. At the next stage Dr. Eve could develop by

- Using Facial Expression Recognition (FER) technology to improve her communication ability
- Using image processing technology to improve her diagnostics ability
- Using other diagnostic tools like digital sphygmomanometer, electrocardiogram (ECG) and electroencephalogram (EEG) to improve her diagnostics ability.